\documentstyle[sprocl,epsfig]{article}

\newcommand{\be}{\begin{equation}}
\newcommand{\ee}{\end{equation}}
\newcommand{\bel}[1]{\begin{equation}\label{#1}}
\newcommand{\bea}{\begin{eqnarray}}
\newcommand{\eea}{\end{eqnarray}}
\newcommand{\beal}[1]{\begin{eqnarray}\label{#1}}

\begin{document}

\title{$J/\psi~c\bar c$ production in $e^+e^-$ and hadronic interactions}

\author{A. B. KAIDALOV}

\address{Institute of Theoretical and Experimental Physics, B.Cheremushkinskaya
25, \\ Moscow 117259, RUSSIA\\E-mail: kaidalov@heron.itep.ru}


\maketitle
\abstracts{ Predictions of the nonperturbative Quark Gluon Strings model,
 based on the 1/N-expansion in QCD and string picture of interactions for
 production of states containing heavy quarks are considered. Relations
 between fragmentation functions for different states are used to predict
 the fragmentation function of $c$- quark to $J/\psi$-mesons. The resulting
 cross section for $J/\psi$-production in $e^+e^-$-annihilation is in a
 good agreement with recent Belle result. It is argued that associated production
 of $c\bar c$ states with open charm should give a substantial contribution to
 production of these states in hadronic interactions at very high energies.}

Investigation of heavy quarkonia production at high energies
provides an important information on QCD dynamics in an
interesting region of intermediate distances from $1/m_Q$ to
$r_{Q\bar Q}$, where $m_Q$ is the heavy quark mass and $r_{Q\bar
Q}$ is the radius of a heavy quarkonia state. For c and b-quarks
this is the region $0.05 fm < r < 1 fm $. In this region both
perturbative and nonperturbative effects can be important.
Production of $J/\psi$-mesons is studied experimentally in
$e^+e^-$ -annihilation,$\gamma p, hp, hA$ and $AA$-collisions.
Analysis of hadronic interactions show that the simplest
perturbative approach (color singlet model) \cite{singlet} does
not reproduce experimental data \cite{review}. This observation
lead to an introduction of the color octet mechanism \cite{octet}
of heavy quarkonia production. In this approach a set of
nonperturbative matrix elements is introduced, which is
determined from a fit to data. A characteristic prediction of
this approach is a large transverse polarisation of $J/\psi$ and
$\psi^{\prime}$ at large transverse momenta \cite{pol} is not
supported by the Tevatron data \cite{Tevpol}.

A new mystery to the problem of heavy quarkonia production has
added recent result of Belle Collaboration \cite{Belle1} on a
large production of $J/\psi$-mesons with charmed hadrons. The
observed cross section at $\sqrt{s}=10.6~GeV$ is an order of
magnitude larger than theoretical predictions \cite{Lieb}, based
on perturbative QCD. It is interesting that at this energy an
associated production of $J/\psi$ with $c\bar c$-pair is the
dominant mechanism of $J/\psi$ production \cite{Belle1}.

In this paper a nonperturbative approach, based on 1/N-expansion
in QCD and string picture of particle production is used for a
description of heavy quarkonia production at high energies. The
model based on this approach (the Quark Gluon Strings model
(QGSM)~\cite{Kaidrev}) has been successfully applied to production
of different hadrons at high energies. It has been also used for
description of inclusive spectra of hadrons containing heavy
(c,b) and light quarks \cite{charm,Pisk,Arak}. In QGSM the
fragmentation functions, which describe transitions of strings to
hadrons in many cases can be predicted theoretically
\cite{Kaidrev,fragm} and are expressed in terms of intercepts of
corresponding Regge trajectories. We will show that the model
naturally leads to the cross section of $J/\psi$ production in
$e^+e^-$ annihilation consistent with the Belle result. Estimate
of the contribution of the same mechanism in hadronic interactions
indicates that it can be important at energies $\sqrt{s}\ge
10^2~GeV$.

Let us first discuss heavy quarkonia production in $e^+e^-$
collisions. In these reactions $c\bar c$-pair is produced
directly by a virtual photon. However a probability of transition
of such a state at high energies (far above threshold of charm
production) to $J/\psi$ is very small. A simplest diagram of QCD
perturbation theory (Fig.1a) corresponds to a transition to a
white $c\bar c$ state with relative momentum characteristic to
$J/\psi$ by emission of two hard gluons. This cross section is
suppressed at high energies by a factor $4m_c^2/s$ and at
$\sqrt{s}=10~GeV$ constitutes ~ $10^{-3}$ of the total $c\bar c$
cross section \cite{Lieb}.

\begin{figure}
\vskip0.05in
\centering
\includegraphics[width=2.5in]{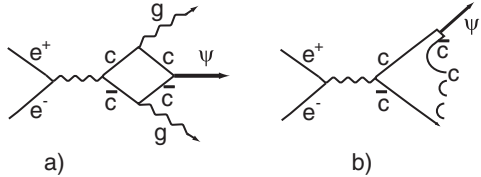}
\vskip 0.1in \caption{ Diagrams for $J/\psi$ production in
$e^+e^-$ annihilation.
\label{fig:e+e-}}
\end{figure}

$J/\psi$ production in association with extra charmed pair
(Fig.1b) does not have this suppression, but contains a smallness
due to production of this pair and a high threshold of the
processes. At high energies this mechanism can be considered as a
fragmentation of $c(\bar c)$ to $J/\psi$. Calculation in  the
lowest order of QCD perturbation theory \cite{Lieb} shows that
this mechanism is important at energies $\sqrt{s}\ge 50~GeV$, but
at $\sqrt{s}= 10~ GeV$ is smaller than the mechanism of Fig.1a by
an order of magnitude and is about 0.07~pb. This is in a sharp
contradiction with Belle result: $\sigma(J/\psi~c\bar
c)=0.87^{+0.21}_{-0.19}\pm 0.17~pb$.

Note that for states of comparatively large radius like $J/\psi$
and especially $\psi^{\prime}$ or $\chi_c$ a nonperturbative
fragmentation can be important. Thus I shall estimate a
fragmentation of $c(\bar c)$ into $J/\psi$ using the
nonperturbative model mentioned above. In this model particle
production is described in terms of production and fragmentation
of quark-gluon strings. A behaviour of the fragmentation
functions is determined in the limit $z\to 1$ from the
corresponding Regge limit and is expressed in terms of Regge
intercepts $\alpha_i(0)$ \cite{Kaidrev,fragm}. The fragmentation
function of $c$-quark to $J/\psi$ in this model is written in the
form \cite{fragm}
 \bel{fragpsi}
D_c^{\psi}=a_{\psi}z^{-\alpha_{\psi}(0)}(1-z)^{-\alpha_{\psi}(0)+\lambda}
\ee where  $\alpha_{\psi(0)}$ is an intercept of the $J/\psi$
Regge trajectory, which is known from analysis of data on spectrum
of $c\bar c$ states and analysis of inclusive spectra of charmed
particles (see below),
$\lambda=2\alpha^{\prime}_{D^*}\overline{p^2_{\perp D}}\approx 1$.
Thus this fragmentation function is characterized by one constant
$a_{\psi}$. In order to determine this constant we will use a
relation between fragmentation function of $c$-quark to $J/\psi$
and fragmentation function of a light quark to $D(D^*)$-meson in
the limit $z\to 1$. According to rules formulated in
refs.\cite{fragm,Bor1} both functions have the same behavior on z:
$(1-z)^{(-\alpha_{\psi}(0)+\lambda)}$ as $z\to 1$ and differ only
by a kinematic factor related to mass difference between $J/\psi$
and $D(D^*)$-meson \bel{ratio}
R^{D/\psi}\equiv\frac{D_u^D}{D_c^{\psi}}=
\left(\frac{s_{0D}^{uc}}{s_{0\psi}^{uc}}\right)^{2(1-\alpha_{D^*}(0))}
\ee The quantities $s_{0i}$ will be determined below.

Now we shall find the fragmentation function $D_u^D$ in the limit
$z\to 1$. In this limit it is related to the fragmentation
function of a light quark to $\pi$ meson \cite{fragm}
$$ R^{D^+/\pi^+}_u\equiv\frac{D_u^{D^+}}{D_u^{\pi^+}}=
\frac{\Gamma^2(1-\alpha_D^*(0))}{\Gamma^2(1-\alpha_{\rho}(0))}
\left(\frac{s_0^{uc}}{\overline{m^2_{D\perp}}}\right)^{2(1-\alpha_{D^*}(0))}
\left(\frac{\overline{m^2_{\pi\perp}}}{s_0^{uu}}\right)^{2(1-\alpha_{\rho}(0))}$$
\bel{ratio1}
(1-z)^{2(\alpha_{\rho}(0)-\alpha_{D^*}(0))}
 \ee
  where $\alpha_{\rho},\alpha_{D^*}(0)$ are intercepts of $\rho$ and
$D^*$  Regge trajectories. They are related to $\alpha_{\psi}(0)$
by the following equation \cite{Kaid1} \bel{trajec}
\alpha_{\rho}(0)+\alpha_{\psi}(0)=2\alpha_{D^*}(0) \ee
 I shall use
the following values for these intercepts: $\alpha_{\rho}(0)=0.5,
\alpha_{\psi}(0)=-2$ and $\alpha_{D^*}(0)=-0.75$ in accord with
 eq.(\ref{trajec}). An uncertainty in the value of
$\alpha_{\psi}(0)$ discussed in
 ref.~\cite{charm} is eliminated at present by experimental data on
inclusive spectra of charmed hadrons in hadronic collisions.

The gamma functions in eq.(\ref{ratio1}) appear from Regge
residues of the corresponding trajectories, which were chosen in
accord with dual models are in a good agreement with data on
widths of hadronic resonances \cite{Volkov}. The coupling is
assumed to be universal (with an account of SU(4) and heavy quark
symmetry).

The quantities $s_{0i}$ entering in eq.(\ref{ratio1}) can be
easily calculated using formulas and parameters of
ref.~\cite{Bor1} \bea
(s_0^{uc})^{2\alpha_{D^*}(0)}=(s_0^{uu})^{\alpha_{\rho}(0)}(s_0^{D\bar
D})^{\alpha_{\psi}(0)};\\
 s_0^{uu}=4m_{u\perp}^2=1~GeV^2 ;~s_0^{D\bar D}=
 (m_{c\perp}+m_{u\perp})^2
\eea
 With $m_{u\perp}=0.5~GeV$ and $m_{c\perp}=1.6~GeV$ \cite{Bor1} we
 obtain $(s_0^{uc})=3.57~GeV$.
Using these values for $s_{0i}$ in eq.(\ref{ratio1}) and
$\overline{m_{\pi\perp}^2}=0.18~GeV^2,~~\overline{m_{D\perp}^2}=5~GeV^2$,
 the fragmentation function $D_u^{\pi^+}=0.44$~\cite{Kaidrev} we
obtain the function $D_u^{D^+}$ at $z\to 1$ in the form
$0.01(1-z)^{(-\alpha_{\psi}(0)+\lambda)}$. This value is in a
reasonable agreement with phenomenological studies of charmed
particle production in hadronic interactions in the framework of
QGSM~\cite{Pisk,Arak}.

   The value of $s_{0\psi}^{uc}$ in eq.(\ref{ratio}) can be
calculated in the same way with the substitution $s_0^{D\bar
D}\to s_0^{\psi D}=6.72~GeV^2$. Finally we obtain from
eq.(\ref{ratio}) \bel{Dpsi}
D_c^{\psi}=0.05~(1-z)^{(-\alpha_{\psi}(0)+\lambda)};~z\to 1 \ee
Thus $a_{\psi}=0.05$.

At asymptotic energies $s\to \infty$ cross section for $J/\psi$
production in $e^+e^-$ annihilation is equal to
\bel{cross1}
\sigma_{\psi}=2~\sigma_{c\bar c}\int_0^1 D_c^{\psi}(z)\,dz
\ee
factor 2 in eq.(\ref{cross1}) takes into account $J/\psi$
production by both $c$ and $\bar c$ quarks. At energy
$\sqrt{s}\sim 10~GeV$ there is an extra suppression due to phase
 space corrections for production of a heavy state. We estimate
 it by introducing an extra factor
 $\gamma=\sqrt{1-4M_D^2/M_{c\bar c}^2}$ to eq.(\ref{cross1}).
 Distribution in $M_{c\bar c}^2$ is related to the z distribution.
 It has a maximum at
 $M^2\approx 0.27s$. For energy of Belle experiment the correction
 factor $\gamma=0.7$. Thus we obtain the following cross section for
$J/\psi~c\bar c$ production at $\sqrt{s}=10.6~GeV ~
\sigma_{\psi}=1.2~pb $. This value is in a good agreement with
Belle  result \cite{Belle1} and is much larger than perturbative
QCD prediction \cite{Lieb}. An estimated uncertainty in the value
of cross section due to possible variation of quantities
$s_{0i},m_{i\perp}$ and ${\alpha_i(0)}$ is about 50$\%$.

\begin{figure}
\vskip0.05in \centering
\includegraphics[width=2.5in]{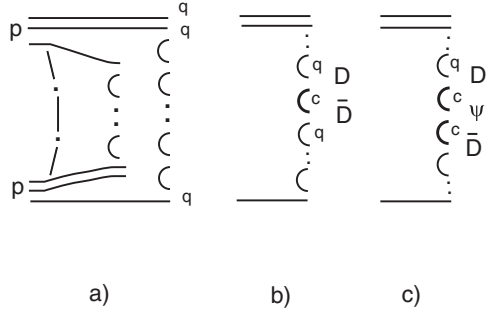}
\vskip 0.1in \caption{ Diagrams for $J/\psi$ production in $pp$-
interactions. \label{fig:pp}}
\end{figure}

Let us consider now $J/\psi$ production in hadronic interactions.
In the approach based on $1/N$-expansion \cite{Hooft} the main
diagrams for particle production correspond to two-chain
configurations, shown for pp-interactions in Fig.2a
\cite{Kaidrev}. They can be considered as production and
fragmentation of two $q-qq$ strings. It is important to emphasize
that production of one $c\bar c$-pair together with light quark
pairs in this approach always leads to an open charm production
(Fig.2b) and  $J/\psi$ in this case is produced by OZI forbidden
mechanism \cite{Bor2}. This leads to a strong suppression ($\sim
10^{-2}$) for heavy quarkonia production in hadronic collisions
compared to open charm (beauty) production. To produce $J/\psi$ in
the chains by OZI allowed mechanism it is necessary to produce
2~$c\bar c$ pairs close in rapidity (Fig2.c). Though this
mechanism is suppressed due to production of extra heavy quark
pair it can compete at very high energy with the mechanism of
single $c\bar c$ pair production. Its contribution can be
estimated from charm quark fragmentation into heavy quarkonia in
$e^+e^-$- annihilation. Consider production of a $c\bar c$-pair in
$q-qq$ string of Fig.2. In each of $q-\bar c$ and $c-qq$
substrings an extra $c\bar c$ pair can be produced and fragment to
a given quarkonium state. So it is possible to use an estimate of
the fragmentation function of $c(\bar c)$ quarks given above or
directly experimental data from $e^+e^-$ to determine a
contribution of the corresponding diagrams to quarkonia
production. This calculation is rather straightforward except of a
threshold suppression factor. It is clear that at energies of
fixed target experiments $\sqrt{s}=10\div 40~GeV$ there is a
strong suppression for production of $J/\psi$ and extra $D\bar D$
pair. I shall estimate this suppression factor for an energy of
HERA-B experiment \cite{HERAB} $E_{Lab}=920~GeV$. Let us denote an
extra suppression factor compared to suppression of a single
$c\bar c$ pair by $\gamma_{pp}$. For its estimation it is possible
to introduce the same kinematical factors as in $e^+e^-$
collisions for each subchains $q-\bar c$ and $c-qq$. For $J/\psi$
production at rapidity y=0 $\gamma_{pp}\approx 0.5$. Another
estimate can be done by assuming that $J/\psi$- $D\bar D$ system
is produced by a gluon fusion. This gives  $\gamma_{pp}\approx
0.4$. Using these estimates and taking into account that
$\sigma_{pp}^{\psi} /\sigma_{pp}^{c\bar c}\approx 10^{-2}$ we
obtain that associated production of $J/\psi$ with charmed hadrons
constitute at this energy $\sim 10\%$. At Tevatron energies the
role of this mechanism is more important and it can (at least
partly) explain an excess of $J/\psi$ production at Tevatron
compared to color singlet model.

For $\psi^{\prime}$ associated production with $c\bar c$ in
$e^+e^-$ annihilation is not known experimentally yet. However
its total inclusive yield is close to the one for $J/\psi$
\cite{Belle2}. If a probabilty of  $\psi^{\prime}$ production by
c-quark fragmentation is the same as for $J/\psi$ it will have
even stronger impact on  $\psi^{\prime}$ production in hadronic
collision because experimentally for $\psi^{\prime}$ cross
section is smaller than for $J/\psi$:
$\sigma_{pp}^{\psi^{\prime}}/\sigma_{pp}^{c\bar c}\approx
1.6~10^{-3}$ and associated production can constitute a large
fraction of the $\psi^{\prime}$ production.

In conclusion it was demonstrated that the nonperturbative QGSM
model predicts a sizable $J/\psi ~c\bar c$- production in $e^+e^-$
annihilation at high energies consistent with recent experimental
result \cite{Belle1}. In the approach based on $1/N$-expansion in
QCD it was shown that a large fraction of $c\bar c$-quarkonia
production in hadronic collisions at very high energies can be due
to associated production with charmed hadrons

\section*{Acknowledgments}

 I would like to thank K. Boreskov, O.V. Kancheli for useful duscussions.
 I am especially grateful to M.V. Danilov for drawing my attention to this
problem and discussion of results of Belle Collaboration. \\
  This work is supported in part by the grants: INTAS 00-00366, NATO PSTCLG-
977275, RFBR 00-15-96786, 01-02-17383

\end{document}